\documentclass{LT23auth}
\usepackage{graphicx}

\begin{document}

\begin{frontmatter}

\title{Electronic structure and the Fermi surface of UTGa$_{5}$ \\
(T=Fe, Co, Rh)}

\author[address1]{Takahiro Maehira \thanksref{thank1}},
\author[address2]{Masahiko Higuchi},
\author[address3]{Akira Hasegawa}

\address[address1]{Advanced Science Research Center, Japan Atomic 
Energy Research Institute, Tokai-mura, Ibaraki 319-1195, Japan}

\address[address2]{Department of Physics, Shinshu University, 
Matsumoto, Nagano 390-8621, Japan}

\address[address3]{Niigata University, Niigata, Niigata 950-2181, Japan}

\thanks[thank1]{ E-mail:maehira@popsvr.tokai.jaeri.go.jp}

\begin{abstract}

The relativistic energy-band calculations have been 
carried out for ${\rm UFeGa_{5}}$, ${\rm UCoGa_{5}}$ and 
${\rm URhGa_{5}}$ under the assumption that 5$f$-electrons are 
itinerant. A hybridization between the U 5$f$ state and Ga 4$p$ state occurs 
in the vicinity of the Fermi level. 
The Fermi surface of ${\rm UCoGa_{5}}$ is quite similar to that of 
${\rm URhGa_{5}}$, which are all small in size and closed in topology.
${\rm UFeGa_{5}}$ has the quasi-two-dimensional Fermi surface which looks
like a lattice structure.

\end{abstract}

%
%
\end{frontmatter}


Uranium compounds exhibit a great variety of phenomena such as Pauli paramagnetism, quadrupolar and magnetic ordering, heavy fermion, and unconventional superconductivity.
It is considered that these behaviours are closely related to the crystal structure, correlation among $f$-electrons, and the $f$-electron hybridization with conduction and/or other valence electrons.
To understand the electronic properties of these materials, the features of the ground state should be first clarified. For this purpose, it is quite important to investigate the Fermi surface both experimentally and theoretically.

${\rm UTGa_{5}}$(T:transition atom) belongs to the heavy fermion compounds with the ${\rm HoCoGa_{5}}$-type tetragonal crystal structure. In order to calculate the energy band structure of uranium compounds, relativity should be take into account because of large atomic number. In this paper, the energy band structures for ${\rm UFeGa_{5}}$, ${\rm UCoGa_{5}}$ and ${\rm URhGa_{5}}$ are calculated by the relativistic linear augmented-plane-wave(RLAPW) method\cite{higuchi}. In the RLAPW method, the four-component relativistic plane wave is augmented by a linear combination of the solutions for the spherically symmetrical potential. The exchange and correlation potential is treated in the local density approximation(LDA). The spatial shape of the one-electron potential is determined in the muffin-tin approximation. The self-consistent calculation is carried out for the experimental lattice constant.

\begin{figure}[btp]
\begin{center}\leavevmode
\includegraphics[width=0.8\linewidth]{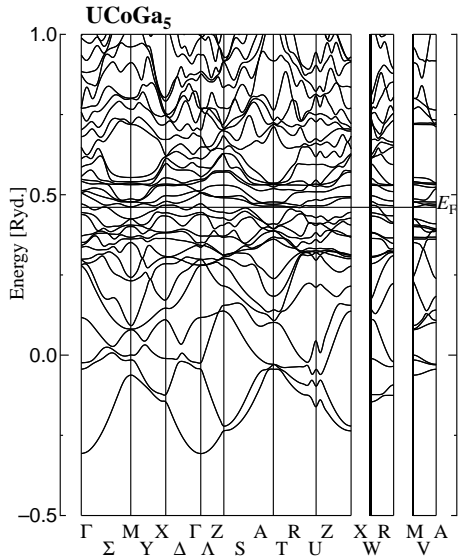}
\caption{ 
Energy band structure calculated for ${\rm UCoGa_{5}}$. $E_{\rm F}$
 shows the Fermi energy.
}\label{ek}
\end{center}
\end{figure}

\begin{figure}[btp]
\begin{center}\leavevmode
\includegraphics[width=0.73\linewidth]{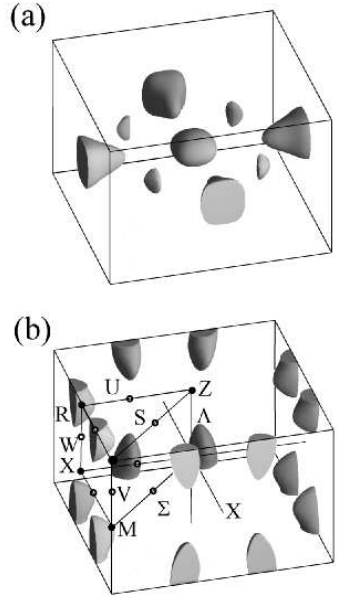}
\caption{ 
The Fermi surface for ${\rm UCoGa_{5}}$ centered at the $\Gamma$ point in the Brillouin zone. (a)The small hole sheet of the Fermi surface in the 15th band. (b)The small electron sheet of the Fermi surface in the 16th band.
}\label{fermi}
\end{center}
\end{figure}

 In Fig.~\ref{ek}, the energy band structure calculated for ${\rm UCoGa_{5}}$ is shown. The Fermi energy $E_{\rm F}$ is located at 0.460 Ryd. Narrow bands, which lie just above $E_{\rm F}$ and are split into two subbands by the spin-orbit interaction, are identified as the U 5$f$ bands. A hybridization between the U 5$f$ state and Ga 4$p$ state occurs in the vicinity of $E_{\rm F}$.
The 15th and 16th bands form the Fermi surface, which is shown in Fig.~\ref{fermi}. These sheets of the Fermi surface are all small in size and closed in topology. 
The hole sheets of the Fermi surface, as shown in Fig.~\ref{fermi}(a), consists of the sheet centered at the $\Gamma$ point, two equivalent sheets centered at the X points and the sheet which lies across the $\Sigma$ axis.
Fig.~\ref{fermi}(b) shows a set of the sixteen electron sheets of the Fermi surface in the 16th band.  Each electron sheet lies across the T axis and looks like a cushions.
The total number of holes is equal to that of electrons, which means that ${\rm UCoGa_{5}}$ is a compensated metal.

The theoretical electronic specific-heat coefficient $\gamma_{band}$ is 4.2 mJ/K$^2 \cdot$mol. It is about the half of the experimental value of 7 mJ/K$^2 \cdot$mol~\cite{okuda}.
This disagreement between $\gamma_{band}$ and $\gamma_{exp}$ should be ascribed to the effects of the electron correlation which the LDA fails to take into account.

\begin{figure}[btp]
\begin{center}\leavevmode
\includegraphics[width=0.75\linewidth]{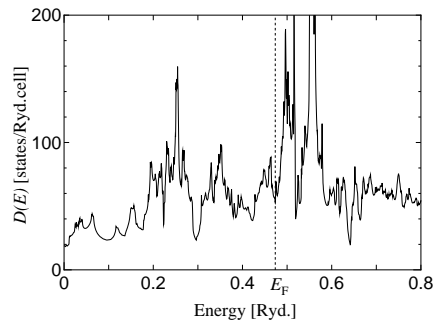}
\caption{ 
Total density of states for ${\rm URhGa_{5}}$. $E_{\rm F}$
 shows the Fermi energy.
}\label{dos}
\end{center}
\end{figure}

Similarly, we calculated the energy band structure of ${\rm URhGa_{5}}$ with the RLAPW method in the LDA.
The 15th and 16th bands construct the Fermi surface, which are very similar to those of ${\rm UCoGa_{5}}$.
Fig.~\ref{dos} shows the total density of states as a function of energy in the vicinity of $E_{\rm F}$.

Finally, the energy band structure of ${\rm UFeGa_{5}}$ is discussed.
The 14th and 15th bands construct the Fermi surface sheets.
${\rm UFeGa_{5}}$ has the odd number of electrons per primitive cell, and therefore becomes the uncompensated metal where the numbers of electrons and holes on the fermi surface are not equal to each other. 
The electron sheets of the Fermi surface in the 15th band are cylindrical, which look like the lattice structure.
The origins of the de Haas-van Alphen frequency branches can be clarified satisfactorily well by our theoretical fermi surface model\cite{tokiwa}.
Note that the electron sheets are quite analogous to those of ${\rm CeIrIn_{5}}$\cite{haga}. 
Due to the simple two-dimensionality of the Fermi surface,
${\rm CeIrIn_{5}}$ is an ideal example for fundamental study of
superconductivity in rare earth compounds.
At present, we construct a microscopic model
by combining the band calculation and the tight-binding 
method\cite{maehira1,maehira2}.

%
%

%
%

The authors would like to thank Kazuo Ueda, Takashi Hotta and 
Tetsuya Takimoto for useful discussions.


%
%

\end{document}